\documentclass[english,aps,10pt,prb, floatfix, notitlepage, superscriptaddress, longbibliography,  nofootinbib,
reprint]{revtex4-2}

\newif\iftwocol
\twocoltrue

\usepackage{lmodern}
\usepackage{amsmath}
\usepackage{amsfonts}
\usepackage{dsfont}
\usepackage{graphicx}
\usepackage{xcolor}
\usepackage{braket}
\usepackage{mhchem}

\usepackage{mathtools}
\usepackage{bm}

\usepackage{booktabs}

\usepackage{wasysym}

\usepackage[unicode=true,pdfusetitle,
bookmarks=true,bookmarksnumbered=true,bookmarksopen=true,bookmarksopenlevel=2,
breaklinks=false,pdfborder={0 0 0},backref=false,colorlinks=true]{hyperref}
\usepackage{bookmark}

\usepackage[columnwise]{lineno} 

\renewcommand*{\vec}[1]{{\bm{\mathrm{#1}}}}

\providecommand*{\moire}{moir\'e}

\providecommand*{\modelname}{crown}

\newcommand*{\heading}[1]{\belowpdfbookmark{#1}{#1}{\bfseries\textit{#1.---}}\ignorespaces}

\usepackage[normalem]{ulem}

\newlength{\apsfigurewidth}
\def \u {\vec{u}}
\def \R {\vec{R}}
\def \r {\vec{r}}
\newcommand*{\ZZ}{$\mathds{Z}_2$}
\def \IPR {\text{IPR}}
\def \del {\nabla}

\newcommand*{\citeapp}[1]{\cite{supplementary}}

\usepackage{comment}
\iffalse

\else
\excludecomment{gobble}
\fi

\begin{document}
	
	\title{Large symmetry and hierarchical ordering transitions in sliding ferroelectrics}
	
	\author{Benjamin Remez}
	\affiliation{Department of Physics, Yale University, New Haven, Connecticut 06520, USA}
	\affiliation{Joint Quantum Institute, University of Maryland, College Park, MD 20742, USA}
    \author{Moshe Goldstein}
	\affiliation{Raymond and Beverly Sackler School of Physics and Astronomy, Tel Aviv University, Tel Aviv 6997801, Israel}

	\begin{abstract}	
	Van der Waals ``sliding'' ferroelectric bilayers, whose electric polarization is locked to the interlayer alignment, show promise for future non-volatile memory and other nanoelectronic devices. These applications require a fuller understanding of the polarization stability and switching properties, which present models have described in terms of an Ising-like binary polarization. However,  it is a much larger translation symmetry that is broken in the polar state.
	Here we introduce a discrete statistical--mechanical model that emphasizes the effect of this larger symmetry. Through Monte--Carlo numerics we show this model possesses a richer phase diagram, including an intermediate critical phase of algebraically-correlated polarization. A low energy effective theory allows us to connect the ferroelectric--paraelectric transition to the Berezinskii--Kosterlitz--Thouless class, driven by excitations not available in Ising-like models. Our results indicate the need for theoretical models of this ferroelectric system to account for the larger symmetry.
	\end{abstract}
	
	\date{\today}
	
	\maketitle


\heading{Introduction} 
The prediction \cite{Li2017a} and discovery of interfacial ferroelectricity \cite{Wu2021, ViznerStern2024} in two-dimensional van der Waals homobilayers, including hexagonal boron nitride (hBN) \cite{Yasuda2021, ViznerStern2021} and transition metal dichalcogenides (TMDs) such as \ce{WTe2}, \ce{WSe2} and \ce{MoS2} \cite{Fei2018, Xiao2020, Wang2022b, Liu2022, Weston2022, Deb2022} attracted significant attention, in part due to its prospective application as non-volatile memory \cite{Yasuda2024, Yang2024, Bian2024}. In these systems, the two opposite polarization states correspond to two degenerate, non-centrosymmetric relative stackings which minimize the monolayers' adhesion energy. As one state is transformed to the other by the rigid displacement of one layer over the other, this phenomena is also known as ``sliding ferroelectricity''.

Early studies focused on the polar structure of twisted (\moire{}) bilayers, both in experiment \cite{ Zheng2020, ViznerStern2021, Woods2021, Lv2022, Ko2023, Zhang2023, Molino2023, VanWinkle2024} and in theory \cite{BennettRemez2021, Bennett2022, Enaldiev2020, Enaldiev2022, Ramos-Alonso2024}. The \moire{} superlattice manifests itself in a high-contrast pattern of alternating positive and negative vertical polarization domains which is easily measured. 
However, \moire{} bilayers are unattractive for device applications, due to their fixed domain texture, vanishing net polarization, challenging fabrication, and propensity for disorder. 
Technological applications would instead benefit from the strong retentive \cite{Yasuda2024} and  cumulative \cite{Deb2022, Cao2024a} polarization of aligned bilayers, which, in the absence of other defects, should form a single epitaxial bilayer with uniform polarization. 
Utilizing sliding ferroelectricity in practice will require an understanding of the room-temperature stability, endurance, and switching dynamics of this polarization \cite{Fei2018,  Liu2022, Yasuda2021, Wang2024, Yasuda2024}.   

Moreover, sliding ferroelectricity also carries fundamental theoretical interest due to its symmetry: One layer can slide over the other indefinitely and in several directions \cite{Marmolejo-Tejada2022}, and hence it is a (discrete) \emph{translation} symmetry which is broken in the polar state. The appropriate order parameter is thus the two-dimensional interlayer displacement. This presents an unusual scenario where the order parameter space is extensively large, scaling with the system physical size [see also Appendix~\ref{app:TM}]. A salient question arises: how can this order parameter be probed by intensive physical observables with only a small number of possible measurement outcomes, such as a binary polarization?

Here we study a statistical-mechanical model inspired by sliding ferroelectrics that incorporates the large parameter space. The promotion of the degree of freedom from binary to vector sets the stage for a  Berezinskii--Kosterlitz--Thouless (BKT) topological phase transition. This mirrors the ``demotion'' in \ce{PbTiO3}-\ce{SrTiO3} films, where strain pins the ferroelectric order to the thin-film plane \cite{Gomez-Ortiz2022}. Therefore, we bring the larger symmetry to the fore by focusing on \emph{topological} interlayer stackig excitations, which were recently identified and classified \cite{Engelke2023}. In this way we complement the recent work of Tang and Bauer \cite{Tang2023}, who established the role that \emph{continuous} excitations, i.e. phonons, play in renormalizing the energy barrier between the two polarization states, and which they studied within an Ising-like double-well picture. 
We find the large symmetry is broken piecemeal in a hierarchy of multiple phase transitions. As suspected, the electric polarization, possessing low symmetry, will turn out to be an observable of weak discriminative power that cannot alone reveal the full phase diagram.

\begin{figure}
	\centering
	\begin{minipage}{\apsfigurewidth}
		\centering
		\begin{minipage}[t]{\textwidth/2}
			\centering
			\textbf{\small (a)}\hspace{3.5cm}\phantom{.}\\\vspace*{-\baselineskip}
			\hfill
			\includegraphics[height = 0.9cm]{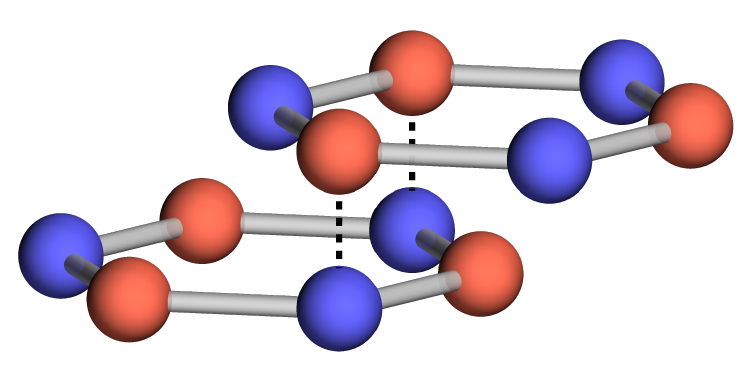}
			\includegraphics[height = 0.9cm]{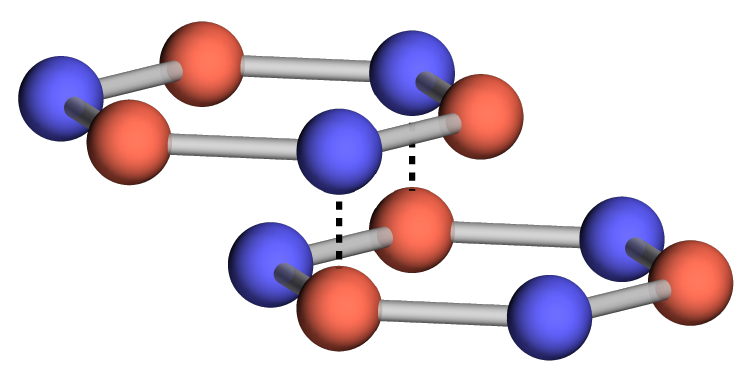}\\
			\includegraphics[width = \textwidth]{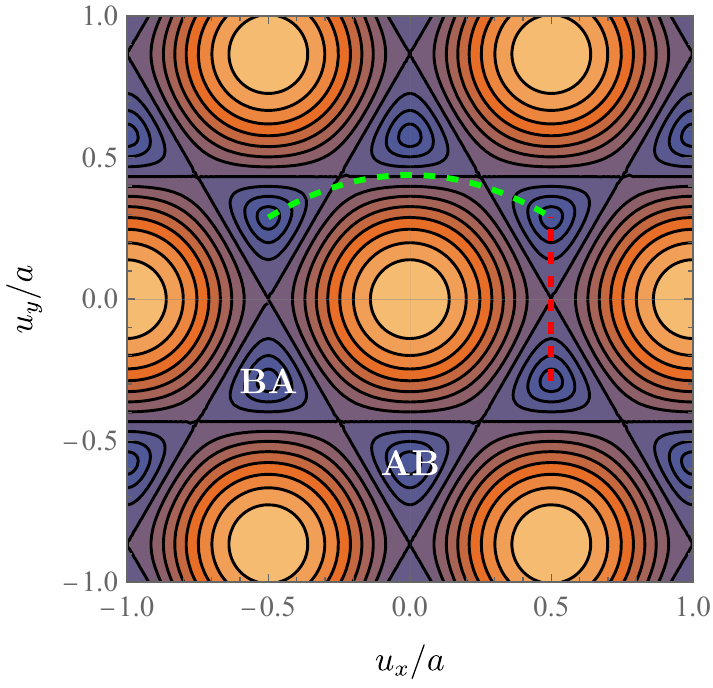}
		\end{minipage}
		\hspace*{-1em}
		\begin{minipage}[t]{\textwidth/2}
			\textbf{\small \,\,\,(b)}\hspace{3cm}\phantom{.}\\\vspace*{-\baselineskip}
			\includegraphics[width = \textwidth, angle=90]{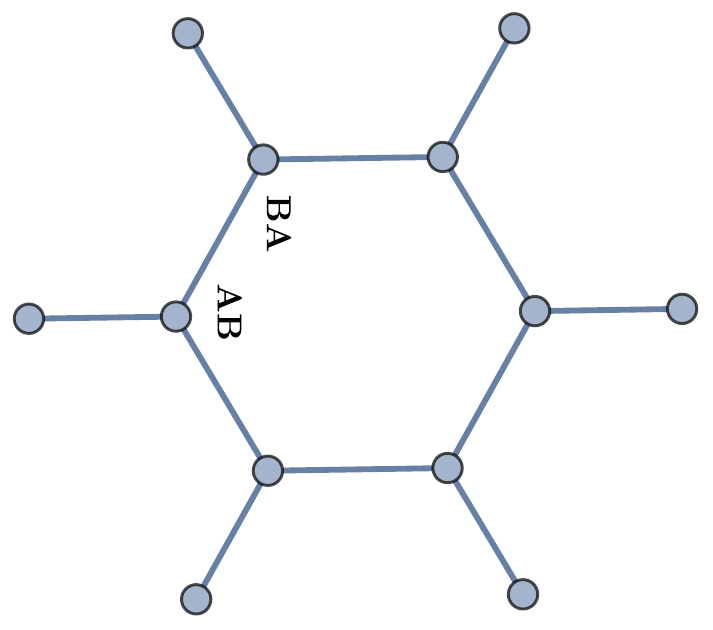}
		\end{minipage}
	\end{minipage}
	\caption{(\textbf{a}) The stacking potential for the interlayer displacement $\u$. 
		The two degenerate minima sublattices correspond to AB and BA bilayer stackings, illustrated above, with red (blue) representing the monolayer A (B) sublattice. 
		The red dashed line represent schematically the trajectory along which $\u$ changes across a domain wall between an AB and BA domains. Two minima on the same sublattice are \emph{not} equivalent, and a domain wall between them (dashed green) costs positive energy. 
		(\textbf{b})~Discrete ``state space''. Vertices correspond to minima of the continuous landscape, and edges indicate the allowed Monte Carlo updates of $\u$. Though the graph is infinite, the low temperature thermodynamics are captured by the finite 12-state subgraph depicted here, which we dub the ``\modelname{}'' model.}
	\label{fig:stacking-potential}
\end{figure}

\heading{Model}
Our work is inspired by hexagonal van der Waals materials such as \ce{hBN}, \ce{WSe2}, etc. (\ce{WTe2} is a notable exception \cite{WTE2}.)  In the bulk, these compounds form Bernal-stacked crystals where successive atomic layers are anti-parallel. Being centrosymmetric, these crystal structures cannot be polar. However, new synthesis methods \cite{Fan2024} like ``tear and stack'' mechanical exfoliation \cite{Kim2016b} and chemical vapor deposition \cite[e.g. in][]{Wang2024, Yang2024} enable the stacking of rhombohedral (R) bilayers of two layers in parallel. Besides alignment, the stacking is specified by the lateral displacement between layers, and R stacking has two degenerate minimal energy configurations known as AB and BA, see Fig.~\ref{fig:stacking-potential}a. Unlike Bernal, these stackings are not centrosymmetric but rather mirror images of each other; therefore, each admits an equal and opposite electric polarization pointing out of the bilayer plane \cite{Ji2023}. The two configurations differ by an offset of the top layer by one-third of a unit cell diagonal. The stacking energy potential for the interlayer offset is depicted in Fig.~\ref{fig:stacking-potential}a.

We model the sliding ferroelectric energy landscape by a discrete elasticity model, 
\begin{equation} \label{eq:model-hamiltonian}
	H = \frac{J}{2} \sum_{\braket{ij}} \left| \u_i-\u_j \right|^2 \,.
\end{equation}
Here $\u_i$ is the local relative atomic displacement between the top and bottom layers. Real space is coarse-grained to sites indexed by $i$, and the sum runs over nearest-neighbor pairs $\langle ij \rangle$. These sites represent the minimal size of a BA domain nucleating in a BA medium (or vice versa), and thus are patches of some mesoscopic size $\ell \gg a$, with $a$ the monolayer lattice constant. We discuss the magnitude of $\ell$ below.  Furthermore, we assume that at low temperatures $\u_i$ only takes values corresponding to the minima of the stacking potential, i.e., it takes values from a honeycomb lattice-like graph, see Fig.~\ref{fig:stacking-potential}b. We call this lattice \emph{state space} to distinguish it from real space. One sublattice of state space corresponds to AB stacking (positive polarization), and the other to BA (negative polarization). To reiterate, in $\u_i$ both the argument $i$ and the value $\u$ take discrete values from a honeycomb lattice. 

Eq.~\eqref{eq:model-hamiltonian} describes the domain wall cost per length $\ell$ between nearest neighbor cells $\braket{ij}$ that realize local stacking configurations $\u_{i,j}$. Microscopically, this domain wall is a crystallographic line defect, and $\u_i - \u_j$ is its Burgers vector. We measure displacement $\u$ in units of the lattice bond length ($a/\sqrt{3}$), so the cost between ``nearest-neighbor'' configurations in state space is $J/2$. Such domain walls could be elastic, as modeled in \moire{} bilayers \cite{Enaldiev2020, BennettRemez2021, Ramos-Alonso2024}, or so-called ``ripplocations'' \cite{Kushima2015} where one layer buckles and folds over the other, as we further discuss below.

Crucially, Eq.~\eqref{eq:model-hamiltonian} admits domain walls between two sites of equal polarization (i.e., $\u_{i} \neq \u_j$ but both on the same state space sublattice), with positive energy cost. This is because microscopically they are separated by multiple domain walls that do not annihilate, as their Burgers vectors do not cancel. For example, a domain wall between ``next-nearest neighbor'' states (Fig.~\ref{fig:stacking-potential}a, green curve) corresponds to a perfect line dislocation, whereas a ``nearest-neighbor'' AB--BA domain wall  (red curve) corresponds to a partial screw dislocation \cite{Enaldiev2022}. We allow arbitrarily large Burgers vectors and use quadratic energy scaling. However, different powers lead to qualitatively similar phase diagrams as the one we demonstrate below (as we have verified numerically), as it is dominated by the values of the first few nearest-neighbor wall costs and not the long range scaling.

Let us now define our observables. The ferroelectric order parameter is 
\begin{equation}
	P = \frac{1}{N}\sum_i P_i = \frac{1}{N}\sum_{i} (-1)^{\u_i} P_0.
\end{equation}
Here $P_i$ is the local polarization dipole, $(-1)^{\u_i}$ is shorthand notation for $+1$ ($-1$) if $\u_i$ is on the A (B) state sublattice, and $N$ is the number of real-space sites. $P_0$ is the magnitude of polarization per unit area in an epitaxial bilayer, of order  $\sim 1~ \mathrm{pC/m}$  \cite{Wu2021a}, which here we set to $1$. Furthermore, we will characterize the phase diagram of model \eqref{eq:model-hamiltonian} by its dipole--dipole correlation function
\begin{equation}
	G(i,j) = G(\left|\R_i - \R_j \right|) = \Braket{P_i P_j},
\end{equation}
where $\left|\R_i - \R_j \right|$ is the real-space distance between sites $i$ and $j$. The averaging is done over a canonical ensemble at temperature $T$ ($k_B = 1$ throughout). Lastly, we define the participation ratio (PR) by its inverse (IPR)
\begin{equation}
	\mathrm{IPR} = \frac{1}{N^2} \big\langle {\sum_{\u} n_\u^2} \big\rangle.
\end{equation}
with $n_\u$ counting how many real-space sites realize the configuration $\u$, and the sum runs over all possible values of $\u$ in state-space.  $1/\mathrm{IPR}$ thus estimates the number of distinct states $\u$ that appear at macroscopic frequency across the real-space lattice.

This system has two limiting behaviors as a function of temperature $T$: (i) At $T \ll J$ we have an epitaxial bilayer and uniform $\u_i=\u_0$ for all $i$ and an arbitrary $\u_0$, so $\IPR{} = P= 1$, and we thus have long range order (LRO) $G(R\to \infty) = 1$. This is the ferroelectric phase. (ii) At $T \gg J$, we expect any positional order to melt, with every possible value of $\u$ realized somewhere, so $\IPR{} = P = 0$ and there is only short range order (SRO) $G(R \apprge 1) \sim e^{-R/\xi}$ with some correlation length $\xi$. This signals a ``liquid'' phase. 
Our focus is on the transition between these limits. Indeed, an intermediate paraelectric solid phase exists below the liquid. Furthermore, we will show that the ferroelectric--paraelectric transition does not occur at a single temperature. Rather, these phases are bridged by a critical fourth phase stretching over a nonvanishing temperature interval. In this critical phase $P=0$ but $\IPR{} \neq 0$, and correlations show quasi-long range order (QLRO), with $G(R) \sim R^{-\eta(T)} $ exhibiting a temperature-dependent critical exponent $\eta$. 

\begin{figure}[t]
	\centering
	\includegraphics[width=\apsfigurewidth]{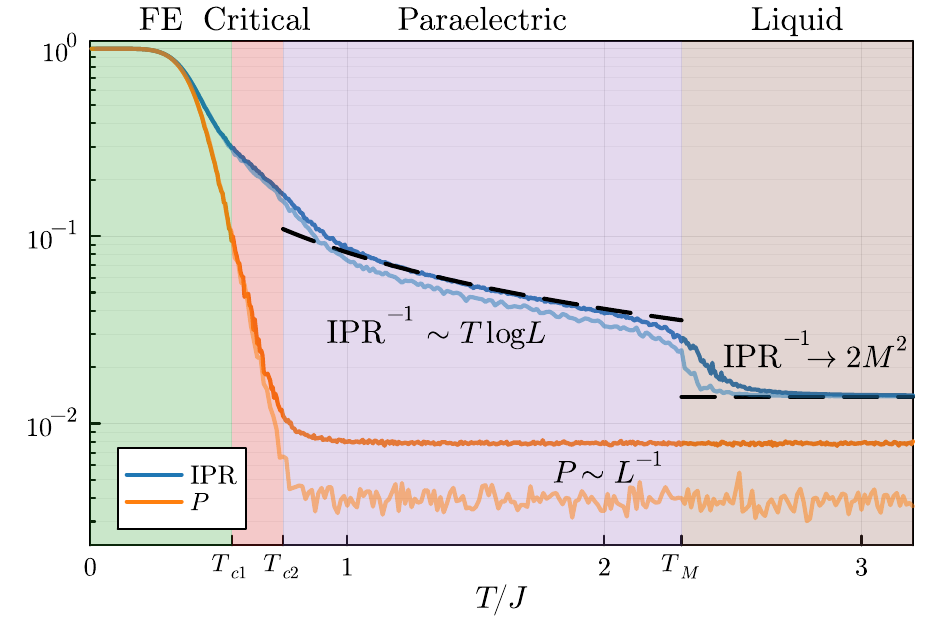}
	\caption{IPR (blue) and polarization (orange) versus temperature, sampled by Monte Carlo.  Opaque (semitransparent) curves correspond to $L=72$ ($L=144$) and $M=6$.  $T_{c1,c2}$ are identified via features in  Figs.~\ref{fig:chi_eta} and \ref{fig:correlation_fits}.}
	\label{fig:ipr_phase_diagram}
\end{figure}

\heading{Monte-Carlo Simulations}
We study Hamiltonian \eqref{eq:model-hamiltonian} by sampling $P$, $G$ and \IPR{} with a Metropolis--Hastings Monte-Carlo simulation \cite{Landau2021}. We take the real space sites $i$ to lie on a honeycomb lattice for implementation simplicity, and as that preserves the $C_3$ symmetry of the constituent monolayers; however, we expect the lattice geometry the play a role subleading to the state space geometry. Unique to the system we study here, we must also truncate the state space lattice to a finite size.  We thus use parallelograms of  $L\times L$ and $M \times M$ honeycomb unit cells for real and state space, respectively, giving $N=2L^2$ real sites and $2M^2$ possible stacking states. 
We implement periodic boundary conditions for both, such that the maximal distance in real and state spaces are $L/\sqrt{3}$ and $M/\sqrt{3}$, respectively. Besides the usual effects of finite real space size, the finiteness and periodicity of state space also lead to interesting, yet spurious effects, which we describe. 

We update all lattice sites $i$ sequentially, ensuring ergodicity by accepting updates with probability at most $1/2$  \cite{Landau2021}. At each time step, configuration $\u_i$ may be updated to one of its three nearest neighbor configurations (see Fig.~\ref{fig:stacking-potential}b). We verify thermalization by annealing from high temperature (defined below \cite{high-TM}) to zero and back to high, confirming we obtain the same sampled traces.

\heading{Phase Diagram}
We first inspect the \IPR{} and polarization $P$ as a function of temperature, which are presented in Fig.~\ref{fig:ipr_phase_diagram}.  

Beginning with low temperatures, below $T_{c1} \approx 0.55\, J$ we find the ferroelectric phase, with $P \ne 0$. Immediately above it is the critical phase, to which we will return shortly.
Subsequently, above $T_{c2} \approx 0.75\, J$, we observe the expected paraelectric phase. In this range the bilayer is depolarized, with $P$ reaching the shot-noise floor $\sqrt{\pi}/L$ and not evolving with temperature. However, the order parameter remains localized to a finite region of configuration space, indicated by a finite $\IPR{} \sim 1/\left(T\log L\right)$. This may be understood as follows: above $T_{c2}$, both real and state space can be taken as continuous, and the Hamiltonian \eqref{eq:model-hamiltonian} becomes a continuum theory 
\begin{equation}
	H_\mathrm{free} = J_\mathrm{free} \int \left(\del u_x\right)^2  + \left(\del u_y\right)^2 d^2 \r \label{eq:H_free}.
\end{equation}
Ergo, we expect $\IPR^{-1} \propto \braket{u^2} \propto T \log L$ by equipartition, with the usual logarithmic divergence of a two-dimensional Gaussian theory. 

Finally, at $ T_M \approx 2.3\,J$ (for $M=6$; see below) the paraelectric phase gives way to complete disorder. The latter is evidenced by the \IPR{} attaining its smallest possible value $1/(2M^2)$, showing $\u$ explores all configurations in state space. 
Fixing $L$ and varying $M$, we see the paraelectric phase extends to higher melting temperatures $T_M \propto M^2$, with negligible quantitative effects on temperatures below $T_M$. We argue in Appendix~\ref{app:TM} that $T_M$ is \emph{not} determined by the saturation condition $\IPR{}^{-1} \sim T\log(L) \approx M^2$. Rather, it represents a BKT transition driven by vortices winding around the toroidal state space, hence $T_M \propto M^2$ with no corrections in $L$. Moreover, in actuality $M\propto L$, hence in the thermodynamic limit $T_M \to \infty$.
We therefore fix $M=6$ hereafter, and focus on temperatures $T < 2J$ \cite{high-TM}.

\begin{figure}[t]
	\centering
	\includegraphics[width=\apsfigurewidth]{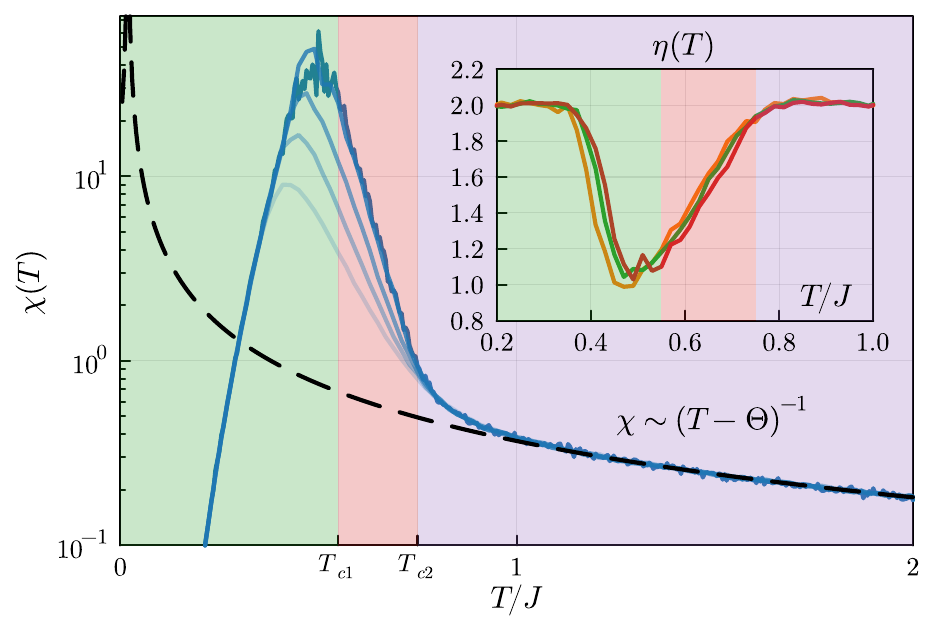}
	\caption{Susceptibility $\chi$ versus temperature. The solid curve (transparent curves) correspond to $L=72$ (increasing $L=8,16,32,64$); $M=6$. The dashed line is a Curie--Weiss fit, see text.
		In the critical phase $\chi$ diverges with system size, $\chi \sim L^{2-\eta(T)}$. The fan-out point at $\approx 0.75\,J$ fixes $T_{c2}$. 
		Inset: Critical exponent $\eta$ obtained by collapsing the curves of the main panel by $\eta = 2-\log (\frac{\chi(L_1)}{\chi(L_2)}) / \log(\frac{L_1}{L_2})$, with $L_2=8$. (SRO yields a formal $\eta=2$ away from the critical phase.) 
	}
	\label{fig:chi_eta}
\end{figure}

\begin{figure}[tb]
	\centering
	\includegraphics[width=\apsfigurewidth]{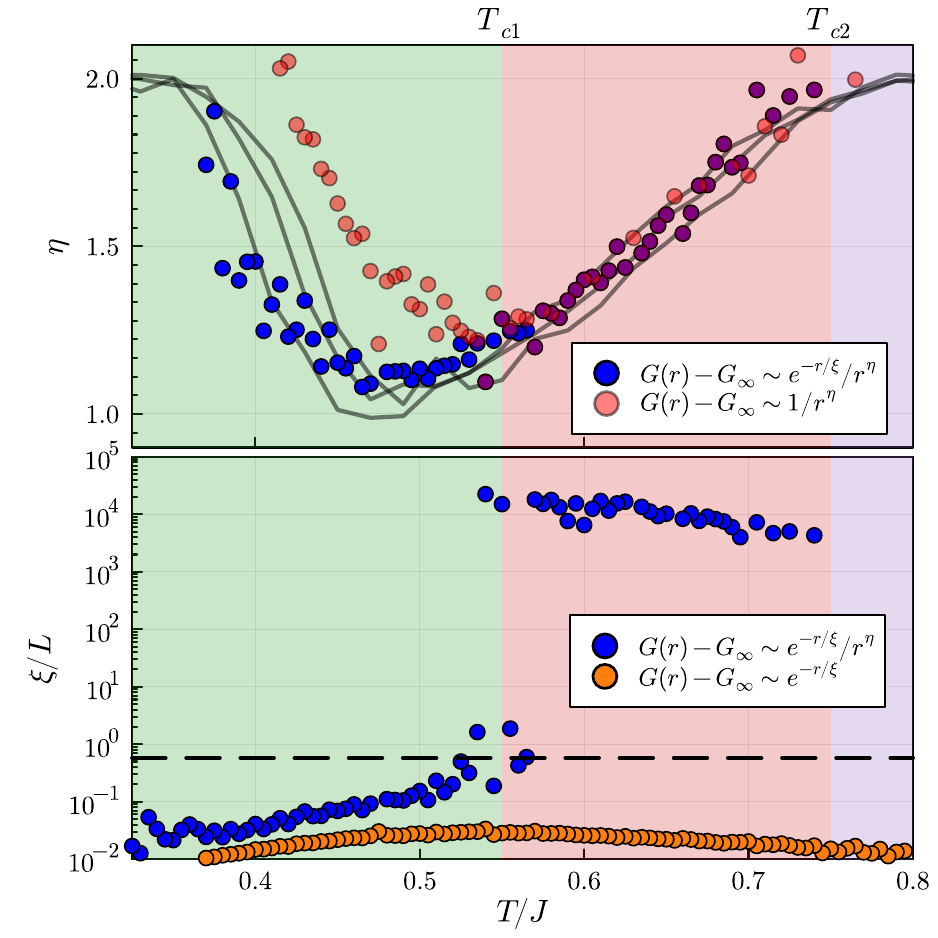}
	\caption{Fitting parameters of the correlation function $G(r)$ for $L=96$, $M=6$. Top: exponent $\eta$ obtained by a fit of a mixed power-exponential dependence $G(r) \sim \exp(-r/\xi)/r^\eta$ (blue) versus a null hypothesis $G(r) \sim 1/r^{\eta}$ (semitransparent red). Over the critical temperature range, the fitted correlation length $\xi \gg L$ by many orders of magnitude (see bottom pane) and the two coincide (appearing purple), demonstrating criticality. The shaded black lines are the same curves from the inset of Fig.~\ref{fig:chi_eta}, showing that exponential corrections well-describe the behavior of $\chi$ just below $T_{c1}$.
		Bottom: correlation length $\xi$ obtained from the mixed dependence fit (blue), versus a null hypothesis $G(r) \sim \exp(-r/\xi)$ (yellow). The dashed line marks the maximal distance $L/\sqrt{3}$, and the bottom of the plot corresponds to one lattice constant. In the ferroelectric phase, power-law corrections reveal the gradual increase of $\xi$ with temperature, reaching up to $L$ and indicating the phase transition at $T_{c1}$. In the critical phase, the fitted $\xi$ is effectively infinite.}
	\label{fig:correlation_fits}
\end{figure}

\heading{Critical Phase}
We now turn our attention back to the interval between $T_{c1}$ and $T_{c2}$.
Over this temperature range, the polar order melts towards zero, whereas the \IPR{} exhibits a ``hump``, indicating bunching of the $\u_i$ in state space. 
We propose that this range hosts an intermediate critical phase. This is first evidenced by the scaling $P\sim L^{-\alpha(T)}$ (not plotted) possessing an exponent $\alpha(T)$ smaller than 1. 
We investigate it further via the dielectric susceptibility, defined as the linear response $\chi = \partial P /\partial E$ to an electric field $E$. $\chi$ is obtained by the usual sampling variance
\begin{equation} \label{eq:chi-variance}
	\chi(T) = \frac{N}{T} \left(\braket{P^2}  - \braket{P}^2 \right),
\end{equation} 
where the factor of $N$ makes it an intensive quantity per unit area. $\chi$ is plotted in Fig.~\ref{fig:chi_eta}. 
In this temperature range we find a pronounced dependence of $\chi$ on the system linear size $L$. This suggests a critical phase with diverging correlation length $\xi$, where algebraic correlations are only truncated by the system size,
\begin{equation}
	\chi \times T = \frac{1}{N}\sum_{i,j} (G(i,j)-\braket{P}^2) \sim \int d^2 \r r^{-\eta} \sim L^{2-\eta},
\end{equation}  
where $\eta$ is the usual correlation function critical exponent. 
This causes all curves to fan out at a single point, which defines our $T_{c2} \approx 0.75 \, J$. By contrast, on approaching $T_{c1}$ from below the curves separate sequentially, as a finite $\xi$ exceeds each value of $L$. We fix $T_{c1}$ below.

By comparing $\chi(T)$ at different system sizes, we  estimate $\eta$, shown in the figure inset, and we find a temperature-dependent $1.0\apprle \eta(T) < 2$ across a finite-width temperature range. This is in contrast to the Ising transition, which occurs at a single critical temperature, at which $\eta = 1/4$ in two dimensions. This clearly demonstrates that despite the binary nature of the polarization, the depolarization transition does not belong to the Ising universality class. 

Additional evidence ruling out an Ising-like behavior can be inferred from the paraelectric susceptibility. Above $T_{c2}$, $\chi$ closely matches a Curie--Weiss law $\chi \sim (T-\Theta)^{-1}$ (This holds in the liquid phase, and $\chi(T)$ is featureless across $T_M$; not shown). The sign and magnitude of $\Theta$ are typically associated with the sign and magnitude of the microsopic ``spin--spin'' interactions.  Here we find a tiny $\Theta \approx 0.02 \, J$. This  suggests that an effective theory with only the local polarization $P_i = \pm 1 $ as the degree of freedom would not be aware of the energy scale $J$, giving yet another evidence of its non-existence. 

Finally, direct evidence of the critical behavior between $T_{c1}$ and $T_{c2}$ can be seen by inspecting the correlation function $G(r)$. In Fig.~\ref{fig:correlation_fits} we show the exponent $\eta$ and correlation length $\xi$ fitted from the form
\begin{equation}
	G(r) = G_\infty +  \frac{A}{r^\eta}e^{-r/\xi},
\end{equation}
where the proportionality constant $A$ and the long-range value $G_\infty$  (typically of order $\braket{P}^2$) are additional fitting parameters. This functional form is general enough to capture all phases, and we benchmark it against pure exponential and power-law fits. We observe that on approaching $T_{c1}$ from below, $\xi(T)$ climbs up to the system size $L$ and then jumps to fitted values many  orders of magnitude larger, indicating a pure power-law dependence. We use this jump to fix $T_{c1}\approx0.55\,J$. 
The fitted values of $\eta(T)$ match those extracted from the finite-size scaling of $\chi$ in Fig.~\ref{fig:chi_eta}, despite the large difference in system sizes. 

The foregoing observations lead us to conclude that the elastic displacements $\u_i$ are fundamental degrees of freedom, which cannot be discarded  in favor of a low-energy effective theory for the local polarization $P_i$.

\heading{Low-temperature model}
The IPR's $L$ dependence is suppressed below $T_{c2}$, see Fig.~\ref{fig:ipr_phase_diagram}. This signals that in this temperature range, $\u_i$ is spread over $\IPR{}^{-1} \apprle 10$ configurations. It is therefore highly localized in state space, indicating that while the \ZZ{} sublattice symmetry is unbroken, state-space translation symmetry is broken. 
Consequently, most of the available state space is not needed to understand the transition at $T_{c1}$. We thus model this low temperature with a reduced model: Let the system be ordered at $T=0$ where all $\u_i$ equal some $\u_0$ somewhere in configuration space. We then consider only a single hexagonal plaquette containing $\u_0$, and the six ``addenda'' configurations that surround that plaquette, see Fig.~\ref{fig:stacking-potential}b, which we call a ``\modelname{}'' model due to its visual shape. Monte Carlo simulations of this restricted space show it agrees with a larger configuration space quantitatively up to $T_{c1}$ and qualitatively up to $T_{c2}$. 

\begin{figure}[t]
	\centering
	\includegraphics[width=\apsfigurewidth]{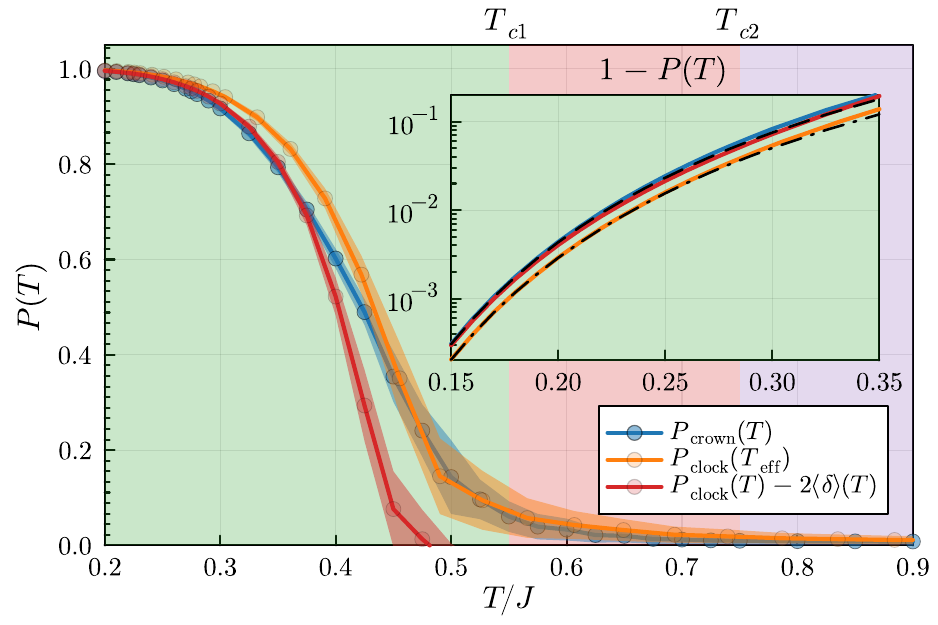}
	\caption{Comparison of the  polarization $P(T)$ in a ``\modelname{}'' proxy to model \eqref{eq:model-hamiltonian} (blue curve) and an effective clock model sampled in Monte Carlo. The latter can be expanded separately around $T \approx T_{c1} $ and $T \to 0 $ (yellow and red curves, respectively). The semitransparent ribbons represent $P$'s thermal variance [i.e. $\sim\chi(T) T$, cf. Eq.~\eqref{eq:chi-variance}] and overlap indicates where the models are in statistical agreement. Inset: low-temperature log-log close-up of the same curves, showing the deviation from order, $1-P$. The dashed and dot-dashed lines are low-temperature cluster expansions for the \modelname{} and clock models [see Appendix~\ref{app:clock}]. }
	\label{fig:EffectiveClock}
\end{figure}

The smaller ``\modelname{}'' model admits a decomposition of $\u_i$ into a modulus $u_i$ and azimuthal angle $\theta_i$. In Appendix ~\ref{app:clock} we integrate out fluctuations of the $u_i$, obtaining at leading order the standard six-sided clock model \cite{Jose1977} for the coordinates $\theta_i$. This angular theory has a renormalized temperature
\begin{equation} \label{eq:Tclock}
	T_\mathrm{clock} (T \apprle T_{c1})  = \left(1 + \frac{1}{1 + e^{\frac{3}{2} J / T}}\right)^{-2} T.
\end{equation}
The fact that $	T_\mathrm{clock}  < T$ indicates that the availability of more configurations surprisingly enhances ordering instead of suppressing it: Fluctuations onto the addenda ring stiffen the model against $\theta_i$ fluctuations, an intriguing example of order-by-disorder \cite{Henley1987} in this system.
However, modulus fluctuations depress the polarization order, since an addenda configuration belongs to the sublattice opposite that of its parent. This kind of dielectric screening similarly renormalizes the polarization
\begin{equation} \label{eq:Pclock}
	P(T \apprle T_{c1}) = \left(1 - 2 e^{-\frac{3}{2} J / T}\right) P_\mathrm{clock}(T_\mathrm{clock}(T)),
\end{equation}  
where $P_\mathrm{clock}(T)$ is the polarization of the clock model. 
This relation has different leading-order expansions at $T=0$ and near $T=T_{c1}$ [Appendix~\ref{app:clock}]. Both are plotted in Fig.~\ref{fig:EffectiveClock}, giving excellent agreement in their respective ranges. 

The clock model exhibits an intermediate critical phase between the ferromagnetic phase with ordered $\theta_i$ and the disordered paramagnetic phases. Rescaling the temperature axis with Eq.~\eqref{eq:Tclock}  allows us to identify the transition at $T_{c1}$ with the ferromagnetic to quasi-long range order transition of the clock model. The critical phase only appears in clock models with at least five ``hour'' positions \cite{Jose1977,Elitzur1979}; our present case of six arose directly from the underlying crystal symmetry of the monolayers. 
From this identification  we infer that in the critical phase $\u$ can have large ``angular'' fluctuations, and therefore would not be captured by a smaller state space model of a central node and its three immediate neighbors.

\heading{Discussion}
Our observations support that the depolarization at $T_{c1}$ is captured by the ordering transition of the clock model. Such identification between the two models loses meaning above $T_{c2}$, where the order parameter becomes deconfined. Yet this does not preclude that this analogy captures the behavior of our model as one approaches $T_{c2}$ from below.  This supports the suggested \cite{ViznerStern2024} connection of the transition at $T_{c2}$ to the clock BKT transition \cite{Berezinskii1972, KosterlitzThouless1973} driven by vortex--anti-vortex unbinding, implying that our model undergoes a topological transition driven by topological defects --- configurations of $\u$ encircling  hexagonal plaquettes in configuration space. It is not immediately obvious that vortex proliferation is compatible, or should be concomitant, with the delocalization of $\u$. However, a far-reaching diffusion of $\u$ in configuration space can indeed be encoded with topological defects; for example, in bilayers with relative twist $\vartheta$,  $\u(\r) \sim \vartheta \hat{\vec{z}} \times \r $ is arbitrarily large far from the twist point, and is realized microscopically by a lattice of vortices forming the \moire{} superlattice \cite{Enaldiev2020, Engelke2023}. We conclude that while the configuration space is very large, the critical properties of the model are captured by the local topology of the configuration space around one plaquette. 

Moreover, on a square spatial lattice and under the Villain approximation, the two clock transitions are dual to each other \cite{Elitzur1979}. This implies, as supported by recent numerical works on the full clock model  \cite{Surungan2019, Li2020, Hostetler2021}, that the transition at $T_{c1}$ is also of the BKT class.  It remains to be verified that these conclusions generalize to models like Eq.~\eqref{eq:model-hamiltonian} with scalings other than quadratic.

The phenomena we describe here occur on the energy and length scales of the model parameters $J$ and $\ell$, respectively. If the domain walls are elastic, these scales are those of a line dislocation in the bilayer. Its width is $\ell \sim a \sqrt{\mu / W}$, with $\mu$ the monolayer shear Lam\'e coefficient, and $W$ the energy barrier between the AB and BA minima (the height of the saddle positions in Fig.~\ref{fig:stacking-potential}). The weak Van der Waals interaction implies $W \ll \mu$, and hence $\ell \gg a$, as we assumed. The dislocation energy per length $\ell$ is $J \sim W\ell\times\ell \sim \mu a^2$. This estimate does not depend on the Van der Waals energy scale, and places the transitions observed here at very high temperatures ($\sim10^4~\mathrm{K}$ for \ce{hBN}), corresponding to the monolayer crystal melting, though consistent with the predictions of Ref.~\cite{Tang2023}, prompting a model that accounts for both mechanisms. However, this estimate is essentially unchanged for \ce{WSe2}, in which the transition has been observed at a much lower $\sim 350~\mathrm{K}$ \cite{Liu2022}. This suggests a more complex interplay of energy scales. 	Indeed, Fig.~\ref{fig:ipr_phase_diagram} is reminiscent of the paradigmatic phase diagram in the KTHNY theory of two-dimensional melting  \cite{KosterlitzThouless1973,Halperin1978, Nelson1979, Young1979}; namely, the paraelectric phase here is analogous to their ``floating solid'' phase of an adsorbate on top of a substrate. However, the floating solid is precluded if the substrate and adsorbate have the same periodicity, as is the case here. The observed \cite{Liu2022} depolarized solid state in \ce{WSe2} indicates a departure from this paradigm. One way in which the mechanics here differ is the availability of an out-of-plane dimension. For example, flextural modes lead to natural rippling in two-dimensional materials \cite{Fasolino2007}. These ripples soften the apparent elastic moduli \cite{Los2016, Los2017}, lowering the relevant energy scales for the sliding physics we discuss here, as compared to the monolayer crystal melting temperature. They may also provide the ``ripplocation'' dislocations that have been suggested as one mechanism for ferroelctric switching \cite{Wu2021}. These membrane-theoretical effects may be important for a quantitative model of this system.

\heading{Experimental consequences}
Our primary observation here is the possibility of phase of quasi-long range polarization order. The ferroelectric order is commonly measured locally by STM \cite{ViznerStern2021, Woods2021}, TEM \cite{Ko2023}, and SNOM \cite{Zhang2023} probes, or globally by transport measurements \cite{Zheng2020, Yasuda2021, Liu2022}. Both approaches are ill-suited for long-range two-point correlation measurements. However, recently developed methods of imaging the polarization texture optically \cite{Kizel2025} could enable such correlation analysis. Let us note that here we do not model the kinetics of the phase transitions; it has recently been proposed that the depolarization results from time-averaging over interlayer sliding events \cite{Marmolejo-Tejada2022}, which may set a temporal resolution constraint to observe these correlations. 
In addition, the slow decay of correlations in real space may hide the nonpolar nature of this phase in small devices, thus masking the intrinsic phase transition.

An externally applied electric field can flip the bilayer polarization by triggering an interlayer slide.
Experiments now demonstrate robust devices withstanding over $10^{11}$ switching cycles \cite{Liu2022, Yasuda2021, Yasuda2024}. Yet a transition between up and down polarization can occur via an interlayer slide in any of the three directions connecting AB and BA minima. Hence, repeated switching events could lead to a ``random walk'' of the interlayer alignment. In practice, this would be mitigated by sample disorder or boundary conditions. However, switching can occur through the growth of nucleating fluctuations; our observation of ``angular fluctuations'' may hint that they will nucleate consistently around the same minima plaquette, providing an intrinsic mechanism that suppresses random walks. This could be investigated in samples comprising a small top flake on a large bottom layer.

\heading{Conclusions and outlook}
In this work we studied the role that an enlarged symmetry may play in the phase transition between two physical states, as motivated by the ferro--paraelectric transition in two-dimensional sliding ferroelectrics.
In fact, we have the unusual situation where the state space scales in size with the real space, which manifests at high temperatures.
Interestingly, at low temperatures leading to the transition, the large symmetry group can distill to an effective smaller group with an endowed relevant topology. In the case of these ferroelectrics, those would be stacking fluctuations around a clock-like model, indicating that the depolarization transition belongs to the BKT class. At temperatures above the transition, the local polarization becomes algebraically correlated.
Further work is required to capture the transition quantitatively.
Some effects which we neglected include the domain wall energy dependence on its orientation \cite{Kushima2015, Enaldiev2022}, the influence of encapsulation in larger heterostructures, and more. 
We emphasize that the model presented here does not pertain to capture the entire behavior of the ferroelectric transition. This includes evidence for a first-order transition \cite{Liu2022}, and that polarization switching takes place by a sweeping domain wall rather than nucleation \cite{Yasuda2024}.

If connection is established between the critical-to-paraelectric transition and the higher-temperature clock transition, this would imply that stacking vortices proliferate in the paraelectric phase. These defects may have a curious non-Abelian algebra \cite{Engelke2023} resulting from vortices surrounding different hexagonal plaquettes in configuration space, making the paraelectric phase an interesting object of study in its own right.


\heading{Acknowledgments}
We thank M.~Ben-Shalom, R.~Fernandes, H.~Ochoa, J.~Junquera, S.~Barraza-Lopez and J.~Roll for fruitful discussions, and D.~Bennett for providing feedback on an early manuscript. B.R.~ gratefully acknowledges the support of the Yale Prize Postdoctoral Fellowship, and the hospitality of the University of Maryland, College Park. M.G. has been supported by the ISF and the Directorate for Defense Research and Development (DDR\&D) Grant No. 3427/21, the ISF grant No. 1113/23, and the US-Israel Binational Science Foundation (BSF) Grant No. 2020072.

	\bibliographystyle{apsrev4-2}
	\bibliography{references,auxiliary_references}

\begin{thebibliography}{55}
\bibitem [{hig()}]{high-TM}%
  \BibitemOpen
  \href@noop {} {}\bibinfo {note} {A finite $T_M$ is convenient for
  initializing the Monte Carlo run. A totally random distribution of $\u_i$ is
  a typical configuration above $T_M$, which we then anneal across $T_M$ to
  reach the colder phases.}\BibitemShut {Stop}%
\end{thebibliography}
	
	\appendix

\section{The transition at $T_M$} \label{app:TM}

One might suppose that $T_M$ signifies the temperature at which the Gaussian fluctuations of $\u_i$ spread over the entire available state space, or $T \log{L}  / J \approx M^2$, suggesting that $T_M \sim J M^2 / \log {L}$ and would vanish in the thermodynamic limit, thus eliminating the paraelectric phase. This is incorrect on two counts: 

(i) The distinct number of interlayer stacking configurations is proportional to the system size, that is, $L / M = \ell/a$ is a fixed finite ratio, and hence the expression above does not tend to zero but in fact diverges with $L$.

(ii) This argument misidentifies the mechanism for the phase transition at $T_M$, which does not follow from a simple saturation of state space. Rather, it is a consequence of the \emph{periodicity} of state space, which (for continuous $\u$) endows it with the topology of a torus $\mathbb{T}^2$ with circumferences $\sim M$. Field configurations $\u(\r)$ thus have point defects classified by their topological invariants in $\pi_1(\mathbb{T}^2) = \mathds{Z}^2$, that is, vortices with winding numbers around the two torus holes. Indeed, $H_\mathrm{free}$ [Eq.~\eqref{eq:H_free}]  represents two identical, decoupled XY ferromagnet Hamiltonians for each component of $\u=(u_x,u_y)$. $T_M$ is thus the BKT transition temperature of the decoupled Hamiltonians, corresponding to $u_{x,y}$ winding around their respective state space boundaries. We have verified that, upon repeating simulations with an elongated state space of dimensions $M_1 \times M_2$, $T_M$ indeed separates into two sequential transitions at temperatures $\propto M_{1,2}^2$, instead of a single one at temperature $\sim M_1\times M_2$.

As is well known, the BKT transition is determined primarily not by single vortex formation energetics but by vortex--anti-vortex interactions, which do not depend on system size. While the IPR as plotted in Fig.~\ref{fig:ipr_phase_diagram} may fulfill the saturation condition at lower and lower temperatures as $L$ increases, the paraelectric phase still has positional QLRO (not plotted), which persists up to $T_M$.

\section{Effective clock model} \label{app:clock}

We compared the phase diagram of Fig.~\ref{fig:ipr_phase_diagram}a with that produced in modified models where the large state-space lattice is pruned down to smaller sizes. No quantitative change was observed below and up to $T_{c2}$  when state space is restricted to 7 hexagonal plaquettes, comprising 24 distinct configuration states, with open boundaries. 

That below $T_{c2}$ we measure $1/\IPR{} \apprle 5$ [cf. Fig.~\ref{fig:ipr_phase_diagram}] indicates that the physics at this range may be described by even fewer sites.
We thus further restrict state space to the 12-site graph depicted in Fig.~\ref{fig:ipr_phase_diagram}b, which we call the ``\modelname{}'' model. 
For clarity, let us designate the 6 states on the inner plaquette as ``core'' states, and the outer 6 states as ``addenda'' states.
We find the \modelname{} model is in quantitative agreement with the full model up to $T_{c1}$, with only small statistically-resolvable deviations in the range between $T_{c1}$ and $T_{c2}$. This agreement highlights the role of the ring-like arrangement of the minima of the interlayer stacking potential.

The austerity of the \modelname{} model, with a single core plaquette, gives rise to the emergent notion of absolute magnitude and angle to the displacement $\u_i$,
\begin{equation}
	\u_i = (1 + R \delta_i) (\hat{\vec{x}} \cos \theta_i + \hat{\vec{y}} \sin \theta_i),
\end{equation}
where $\delta_i = 0,1$ indicates if the configuration $\u_i$ is on the core or the addenda shell, and $\theta_i = 2\pi n_i / 6$, where $n_i = 0,1,\dots5$ is its discrete azimuthal angle. Here $R = 1$ which we retain formally for order-counting.
With this decomposition, Eq.~\eqref{eq:model-hamiltonian} takes the form $
H  = H_\theta + H_\delta + V$, where (in this section we set $J=1$ for brevity)
\begin{align}
	H_\theta  &= \sum_{\braket{ij}} [1-\cos \theta_{ij}]; \\ H_\delta &= \frac{1}{2} R^2  z \sum_i \delta_i; \\
	V &= \sum_{\braket{ij}} \left[R(\delta_i +\delta_j)+R^2 \delta_i \delta_j \right] (1-\cos \theta_{ij}) - R^2\delta_i \delta_j.
\end{align}
Here $z$ is the real-space lattice coordination number (in this work, 3), and $\theta_{ij}=\theta_i-\theta_j$. Here we separated out the decoupled angular and radial dynamics, $H_{\theta,\delta}$, and their coupling expanded in powers of the deviations from ordering, $(1-\cos \theta_{ij})$ and $\delta_i$ (note the term $\propto R$ in $V$ is not included in $H_\delta$).

Our goal now is to describe the low-energy dynamics of $\theta_i$.
$H_\theta$ is the standard Hamiltonian of the ferromagnetic clock model for the effective ``spin'' $\u_i$. Meanwhile, $H_\delta$ indicates that the $\u_i$ prefer to reside on the inner core states, as states on the addenda shell typically have longer distances between each other. However, $V$ implies that $\u_i$ on the high-energy addenda behave as longer spins with stronger ferromagnetic interaction. Therefore, thermal fluctuations of $\delta_i$ will promote and fortify the angular ordering. We may integrate them out by the usual procedure,
\begin{equation}
	H_\mathrm{eff} = H_\theta - T \log \Braket{e^{-\beta V}}_{H_\delta}.
\end{equation}
where $\beta$ is the inverse temperature and the angle brackets denote thermal averaging over the $\delta_i$ with respect to $H_\delta$. To leading order in $V$, this gives
\begin{align}
	& H_\mathrm{eff}  = H_\theta + \sum_{\braket{ij}} (2R \braket{\delta}+R^2 \braket{\delta}^2) (1-\cos \theta_{ij}) + \mathrm{const.} \nonumber\\
	& \simeq (1 + R \braket{\delta})^2 \sum_{\braket{ij}} (1-\cos \theta_{ij}) =  (1 + R \braket{\delta})^2 H_\theta,
\end{align}
where in the second line we dispensed with temperature-dependent constants. The mean radial fluctuation follows a Fermi--Dirac-type distribution
\begin{equation}
	\braket{\delta} = \frac{1}{e^{\frac{1}{2}z\beta R^2} + 1}.
\end{equation}

$H_\mathrm{eff}$ thus has the form of $H_\theta$, but with a renormalized clock temperature temperature, $T_\mathrm{eff} =  (1 + R \braket{\delta})^{-2} T$. Substituting $\braket{\delta}$, $z = 3$, and $R=1$, we arrive at Eq.~\eqref{eq:Tclock}. Hence, at this order in $V$, the thermodynamics of the \modelname{} model are given by that of the six-sided clock model, up to a (nonlinear) rescaling of the temperature axis. This rescaling indeed aligns the transition at $T_{c1}$ of the two models, cf. Fig.~\ref{fig:EffectiveClock}.

Besides the angular order, since each addenda state in the \modelname{} model is on the state-space sublattice opposite to that of its parent in the core, radial fluctuations will also reduce the sublattice polarization order. We may capture this effect by adding a term to the Hamiltonian $H_P = E\sum_i (-1)^{\delta_i}p(\theta_i) $ where $E$ represents an externally applied electric field and $p(\theta) = \cos (3\theta)$ captures the staggered polarization as we go around the \modelname{} plaquette. To lowest order in $E$, integrating out the thermal fluctuations yield
\begin{equation}
	H_{P,\mathrm{eff}} = \Braket{H_P}_{H_\delta} 
	= (1-2 \braket{\delta}) H_P
\end{equation}
where we used $(-1)^\delta = 1 - 2 \delta$ for $\delta=0,1$. Once more, the total rescaling can be reabsorbed in a renormalized external field, $E_\mathrm{eff} = (1-2\braket{\delta}) E$, which signifies a dielectric screening  by the radial fluctuations. Recalling the preceding temperature renormalization, the polarization is
\begin{multline} \label{eq:screening}
	P(T) = (1-2\braket{\delta}) P_\mathrm{clock}(T_\mathrm{eff}) = \\
	=\begin{cases}
		P_\mathrm{clock}(T) - 2 \braket{\delta} + \mathcal{O}[\braket{\delta}^2], & T \to 0^{+}, \\
		P_\mathrm{clock}(T_\mathrm{eff}) + \mathcal{O}[(T_{c1} - T)^b\braket{\delta}], &  T \to T_{c1}^{-}.
	\end{cases}
\end{multline}
Here $P_\mathrm{clock}(T)$ is the polarization of the six-sided clock model (with unit coupling $J=1$) at temperature $T$. In the last line we expanded to leading order: At low temperatures, $1 - P \sim \braket{\delta}$ due to rare fluctuations, whereas near $T_{c1}$ we expect $ P \sim (T_{c1} - T)^b$ with some thermodynamic exponent $b$. These two relations are plotted in Fig.~\ref{fig:EffectiveClock}.

Another perspective for obtaining  Eq.~\eqref{eq:screening} is a cluster expansion. For an ordered system in some uniform state $\vec{u}_0$, we expand the polarization in clusters of up to three sites, obtaining
\begin{equation} \label{eq:cluster}
	P_g(T) = 1 - \frac{2g}{g + e^{3\beta/2}}
	-\frac{6g}{g + e^{4\beta/2}} 
	- \frac{18g}{g + e^{5\beta/2}} + \mathcal{O}(e^{-6\beta/2})
\end{equation}
where the three nontrivial terms arise from the thermal probabilities for one-, two-, and three-site clusters of an opposite domain to occur, and $g$ is the number of state-space neighbors of $\u_0$ (2 for clock, 3 for \modelname{} core states). These clusters have a perimeter of 3, 4, and 5 real-space neighbors, respectively. Here we truncated the expansion by neglecting 4-site clusters (minimal perimeter of 6) as well as any cluster interdependence, both entering at order $\sim e^{-6\beta/2}$. 
We plot $P_{2,3}$ in the inset of Fig.~\ref{fig:EffectiveClock}, showing excellent agreement with numerics up to $T \approx T_{c1}/2$,
and from Eq.~\eqref{eq:cluster} indeed $P_3(T)/(1-2\braket{\delta})P_2(T) = 1 + \mathcal{O}(e^{-4\beta/2})$. 


\end{document}